# Reverse chemistry of iron in the deep Earth

Xiaoli Wang,[1,2,3] Xiaolei Feng,[4,5] Jianfu Li,[1] Dalar Khodagholian,[6] Jiani Lin,[1] Matthew G. Jackson,[2*] Frank J. Spera,[2] Simon A. T. Redfern[4,5*] and Maosheng Miao[6,3*]

[1] *School of physics and electronic engineering, Linyi University, Linyi 276005, People's Republic of China*

[2] *Department of Earth Science, University of California Santa Barbara, Santa Barbara, California 93110, United States*

[3] *Beijing Computational Science Research Center, Beijing 100084, People's Republic of China*

[4] *Department of Earth Sciences, University of Cambridge, Downing Street, Cambridge, CB2 3EQ, UK*

[5] *Center for High Pressure Science and Technology Advanced Research (HPSTAR), Shanghai 201203, China*

[6] *Department of Chemistry and Biochemistry, California State University, Northridge, California 91330, United States*

**Abstract:** In this work, we demonstrate a remarkable change of chemical trend of Iron under high pressure that is of great importance for understanding the distribution of elements in the Earth's mantle and core. Using first principles crystal structure search method, we conduct a systematic study of the propensity of *p*-block elements to chemically bind with iron under high pressures ranging from ambient conditions to that of Earth's core. We show that under increasing pressure, iron tends to reverse its chemical nature, changing from an electron donor (reductant) to an electron acceptor, and oxidizes *p*-block elements in many compounds. Such reverse chemistry has a significant impact on the stoichiometries, bond types and strengths, structures and properties of iron compounds under deep planetary conditions.

# 1. Introduction

The distribution and abundance of both major and trace elements in the Earth's interior provide a record of its formation and evolution [1,2]. An understanding of this record demands knowledge of the chemical affinity of the elements and their compounds under the high-pressure conditions of Earth's interior. For many years, our understanding of such affinities has been predominantly biased by low-pressure observations that are of dubious applicability to Earth's deep mantle and core [3]. Many trace elements are found to have greatly reduced concentrations on Earth relative to their solar abundance [4], [5]. This is usually explained in terms of either the escape of elements to space due to volatility during the high-energy conditions of terrestrial accretion [3], or the incorporation into the Earth's core [6]. The core sequestration model relies on the reactivity of trace elements with Fe (and Ni) under high pressure, which is problematic to assess due to the difficulty of experimentally achieving terrestrial core pressures (135-367 GPa).

Thanks to improvements in computational power and methods, the high-pressure chemistry of Fe has become accessible, leading to the discovery of a number of new Fe compounds with trace elements that supports the argument that they are incorporated in the core. For example, recent work showed that iron may actually bind strongly with xenon to form an $Fe_3Xe$ compound at the pressures of Earth's core, suggesting that core sequestration is the cause of the "missing xenon paradox" [7,8]. A similar mechanism was suggested for the depletion of iodine in Earth [9], although the volatility of the iodine renders this explanation ambiguous. The reactions of Fe with major elements such as O also becomes quite unusual at very high pressure. As revealed by both computer simulation and diamond anvil cell (DAC) experiments, iron can form an oxygen-rich $FeO_2$ compound at the pressures of Earth's lower mantle, even if it remains in the low oxidation state of +2. [10] We show here that these striking phenomena are all related to dramatic changes in 'iron chemistry' under high-pressure. The broad-ranging chemical trends of iron can only be revealed by a large-scale study of iron reactivity across the periodic table, a task that cannot be performed experimentally with reasonable resources and time.

Many recent studies show that first principles structure predictions are sufficiently advanced that enthalpies of compound formation at high pressure can be accurately calculated and the nature of the chemical bond elucidated [7–15]. Using this approach, we have systematically explored the bonding of iron with *p*-block elements in the periodic table. High pressure greatly enhances bonding to iron for many *p*-block elements that are conventionally labeled lithophile or chalcophile [1,16], making them highly siderophile. However, the depletion of the *p*-block elements in silicate Earth correlates inversely with Fe binding strength. This striking result suggests that although the Earth's core can host large quantities of the *p* elements, it is not the cause of their depletion. Instead, cosmochemical accretion models that call on elemental loss by volatility during high-energy conditions of terrestrial accretion may be more relevant [17]. Furthermore, silicon shows a distinct anomaly in its bonding to iron, such that it becomes one of the strongest under high pressure, which suggests silicon may readily be incorporated into Earth's core, corroborating recent perspectives on the composition of Earth's core based on sound speed measurements, experimental petrology and seismology [18,19].

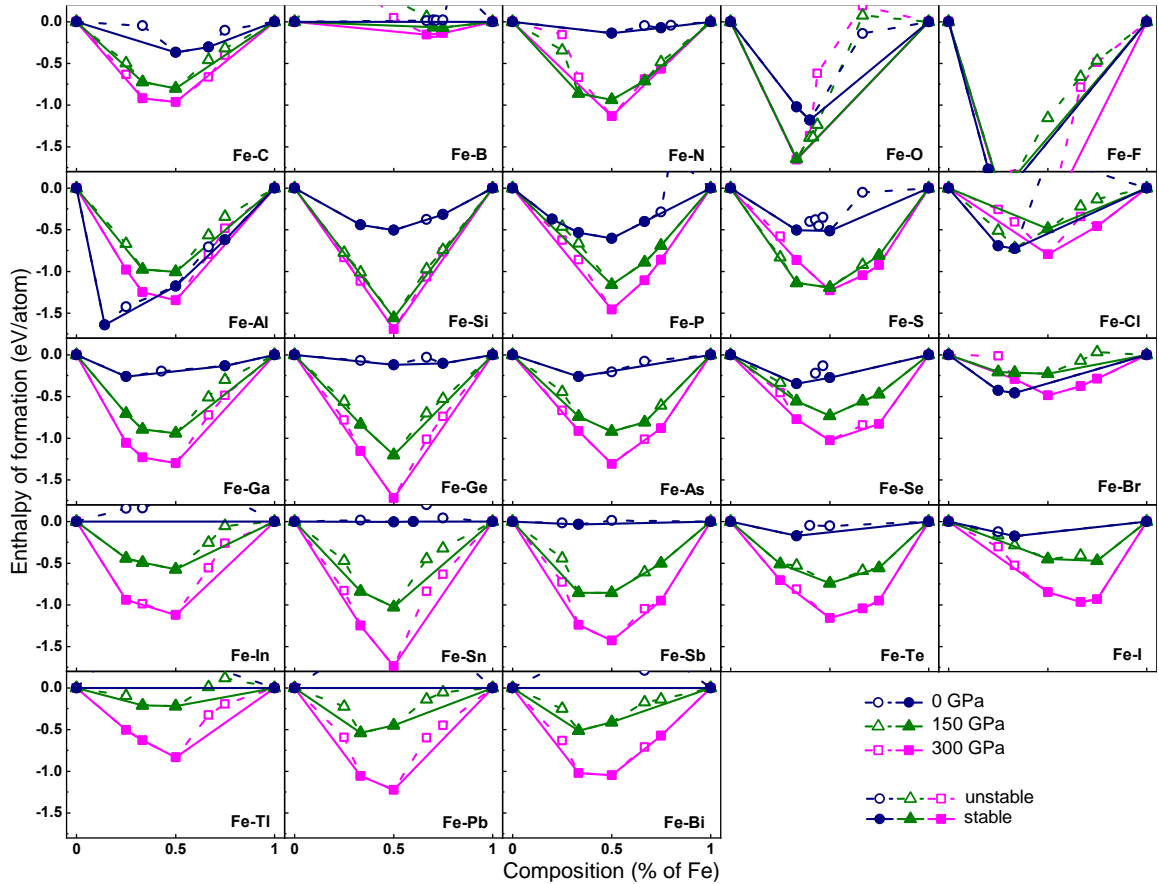

**Fig. 1. Thermodynamic stability of main *p*-block – iron compounds at both ambient and high pressures.** Most structures at ambient pressure are chosen from Materials Project [20]. Most structures of Fe$_m$X$_n$ (*m/n*=1–3) at high pressures are obtained from crystal structure searches using CALYPSO (details are shown in Methods). Convex hulls are shown as solid lines, with stable compounds shown by solid symbols. Unstable compounds (open symbols) sit above convex hulls, with dotted lines indicating possible decomposition routes.

## 2. Computational methods

We performed structure predictions through a global minimization of free energy surfaces based on the CALYPSO (Crystal structure AnaLYsis by Particle Swarm Optimization) methodology as implemented in CALYPSO code [21,22]. We searched the structures of stoichiometric Fe$_m$X$_n$ (*m* = 1−3; n = 1−3) with simulation cell sizes of 1−4 formula units (f. u.) under pressures of 150 GPa and 300 GPa, respectively. All the structures are optimised at a higher accuracy. The calculations for local structural relaxations and electronic properties were performed in the framework of density functional theory within the generalized gradient approximation Perdew-Burke-Ernzerhof (GGA-PBE) [23] and frozen-core all-electron projector-augmented wave (PAW) method [24,25] as implemented in the VASP code [26]. A cutoff energy of 700 eV and appropriate Monkhorst−Pack [27] *k*-mesh with *k*-points density 0.03 Å$^{-1}$ were chosen to ensure that all the enthalpy calculations were well converged to less than 1 meV/atom. For the most stable structures at each pressure, the formation enthalpy per atom is calculated using the following formula:

$$H_f (\text{Fe}_m\text{X}_n) = [H(\text{Fe}_m\text{X}_n) - mH(\text{Fe}) - nH(\text{X})]/(m+n)$$

where $H_f$ is the formation enthalpy per atom and $H$ is the calculated enthalpy per chemical unit for each compound. The enthalpies for Fe and *X* are obtained from the most stable structures as searched by the CALYPSO method at the desired pressures.

## 3. Results and discussion

Our calculations reveal that pressure can dramatically increase the stability of iron compounds formed with most *p*-block elements as evidenced by a significant decrease of formation enthalpy (Fig. 1). This pressure-enhanced Fe reactivity may promote the incorporation of many *p*-block elements, especially the heavier ones that were previously disregarded due to their weak or absent binding with Fe, into Earth's core. Like the previous works, our results first appear to support the model of core sequestration, *i.e.* the depletion of certain elements in the silicate Earth is due to their incorporation into Earth's core. However, the integrated picture that compares the abundance of elements and their binding strength with Fe across the *p*-block of the periodic table shows the opposite trend. The *p*-block element abundances, normalized to CI chondrites, are *inversely* correlated with their binding strength to Fe as quantified by the formation enthalpies of the most stable compounds (Figs. 2a – 2c), *i.e.* the stronger they bind with Fe the less they are depleted in the silicate Earth. While pressure increases, the correlation represented by the shaded stripes becomes steeper due to stronger binding to Fe (more negative enthalpy of formation) for the heavier elements such as Se, Te, Bi, Pb etc. This inverse correlation clearly shows that reaction with Fe in Earth's core is unlikely to be the cause of the depletion of trace *p*-block elements in the silicate Earth, although such reactions may become exceedingly strong under terrestrial core conditions. One reason for the inverse correlations is that the volatility of an element (quantified by its condensation temperature) correlates inversely with its binding strength with Fe, *i.e.* the higher the volatility the weaker the binding with Fe (Fig. S1). Another factor is that binding of an element with Fe may prevent its evaporation in primordial Earth.

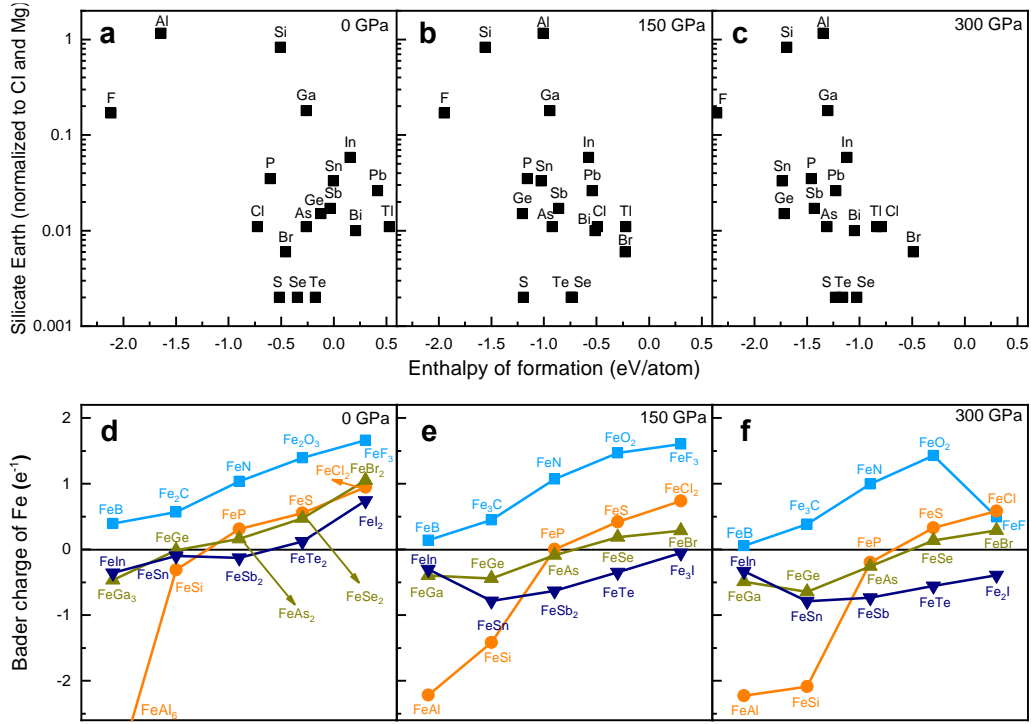

**Fig. 2. Binding strength with Fe, their correlation with depletion in Earth and their chemical origin.** a. – c. The correlations between element depletions and their binding strength with Fe under 0, 150 and 300 GPa. Horizontal axes show the enthalpy of formation per atom of the most stable Fe compound of selected element. The vertical axes show the abundance of an element normalized to CI chondrite and Mg. d. – f. Calculated Bader charge of iron in Fe – $X$ compounds at 0, 150 and 300 GPa.

The dramatic change of Fe chemistry under pressure can be understood by a topological investigation of the electron density, employing Bader charge analysis [28]. The charge transfer between iron and *p*-block elements shows that, apart from Al and Si compounds (see below), the magnitude of charge transfer changes almost linearly with atomic number in each period from group 13 to group 17 (Fig. 2d – 2f). At 0 GPa, the charge is negative for most of the *p*-block elements, indicating that iron is oxidized. The charge transfer is generally small for heavy *p*-block elements, which explains the poor stability of their Fe compounds. Under high pressure, the electrons redistribute toward Fe, reducing its positive charge. For heavy *p*-block elements, like Ge, P, As, Te and I, the charge on Fe changes from positive to negative at 0, 150, 110, 30, and 100 GPa, respectively (Fig. 2d – 2f). Similar charge transfer reversal (CTR) also happens in Fe – B and Fe – Se compounds, but at higher pressures of 375 and 405 GPa, respectively. This CTR changes the chemical character of iron from being a reductant (electron donor) to an oxidant (electron acceptor). CTR is most dramatic for the 5*p* elements such as Te and I. For I, the charge on Fe changes from +0.75*e* at 0 GPa to -0.39*e* at 300 GPa, and CTR happens at 100 GPa. In other words, iron iodide becomes iodine ferride above 100 GPa. The charge redistribution mainly happens between Fe *3d* and X *np* orbitals (Fig. S2), which is the natural result of the energy shifts of the *3d* and the *np* bands (Fig. S3). The Fe *3d* bands become lower in energy because they have a smaller radius and are therefore less prone to change under increasing pressure (Fig. S4).

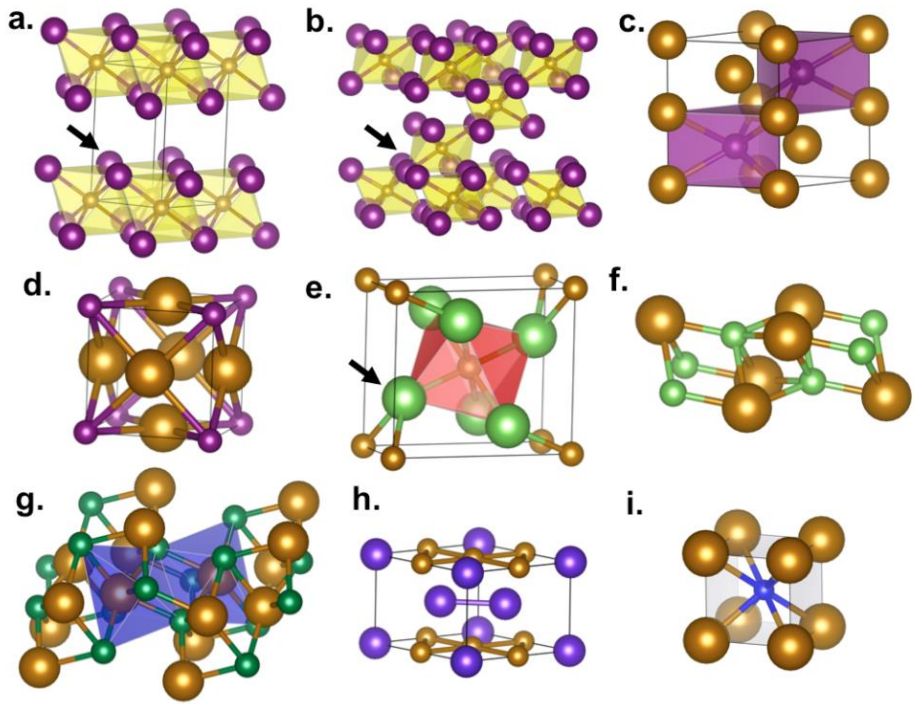

**Fig. 3. Structure evolution of Fe-*p* block compounds under pressure. a.** Layered structure of stable FeI$_2$ at 0 GPa; **b.** FeI$_3$ at 0 GPa; **c**, Fe$_2$I at 150 GPa; **d.** Fe$_3$I at 150 GPa; **e**, Fe*X*$_2$ (*X* = As and Te) structure at ambient pressure; **f.** FeAs at 150 GPa; **g.** FeTe at 150 GPa; **h.** Fe*X* (*X* = Sn and Ge) with intermetallic bonds between Fe atoms; **i.** CsCl structure, a common structure for Fe*X* compounds under pressure. Brown balls represent Fe atoms; the size (small and large) indicates the charge transfer out of or into Fe. The arrows show the interstitial space for lone pair electrons of p block elements. More
7

structures can be found in Fig. S5. The dynamic stability of the structures can be seen in Fig. S6.

Charge redistribution also shows a strong effect on the structure evolution of Fe compounds. At low pressure, many structures contain lone pair electrons on *p*-block elements that will disappear under high pressure, accompanying by the increase of coordination number and charge transfer back to Fe (Figs. 3a – 3d for Fe-I; Figs. 3e – 3g for Fe-As and Fe-Te). Some compounds contain Fe-Fe bonds in the low-pressure structure that simply vanish as pressure increases (Fig. 3h). It is interesting that many FeX compounds adopt the simple CsCl structure under high pressure, due to the large charge transfer to Fe.

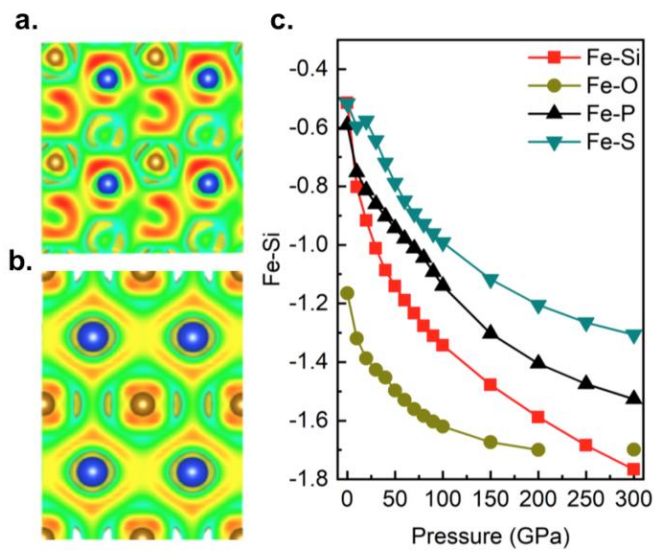

**Fig. 4. Anomaly of Si in binding with Fe.** Electron localization functions (ELF) of FeSi **a**. at 0 GPa and **b.** at 300 GPa; **c.** the enthalpy of formation of Fe-X compounds as functions of pressure. The isosurfaces of the ELFs are taken at 0.2.

Another distinctive feature emerging from the comprehensive computation of Fe chemistry is an anomaly in reacting with Si (and Al). Generally, the charge transfer from iron to *p*-block elements decreases with increasing atomic number in the same group (negative with CTR). For example, the Bader charges on Fe in pnictides increase in the order N>P>As>Sb. Similar increases can be found across the chalcogens and halogens. This order of increasing charge remains under increasing pressure (Fig. 2d – 2f). However, for group 14 elements, the charges of Fe at 0 GPa increase in the order C>Ge>Sn>Si, and those for the last three elements are very close to one another. Under high pressure, the charge of Fe in Fe-Si compounds decreases most dramatically and becomes significantly lower than Ge and Sn. A similar phenomenon can be seen for group 13 elements. This distinctive anomaly is due to the fact that the unoccupied *d* shells are significantly higher in energy for Si (and Al) and cannot host electrons, in contrast to heavier *p*-block elements, leaving large charge transfer from Si (and Al) to Fe, especially under high pressure. At 300 GPa, the charge on iron in FeSi is as low as -2*e*, indicating the very strong ionic nature of the compound. Indeed, the ELF values between Fe and Si decrease dramatically under pressure (Fig. 4a – 4b). Consequently, Fe-Si bonding persistently strengthens at increasing pressure. At 0 GPa, the Fe-Si bond strength is similar to Fe-S and Fe-P, all much weaker than Fe-O (Fig. 4c). Under increasing pressure, the Fe-Si binding strengthens most significantly and even surpasses Fe-O at 250 GPa, implying that Si becomes highly siderophile at the pressure of Earth's core. The lower density of FeSi (Extended Data Fig. 7) under core pressures can also explain the low core density as revealed by the seismic wave measurements. Finally, Ca/Al and Mg/Si ratios are expected to exist in chondritic relative proportions in the silicate Earth, but both ratios are higher than chondritic in the shallower, accessible Earth, suggesting either volatility controlled fractionation of Mg/Si or the presence of an Al or Si-rich domain in Earth's deep interior [29–31]. Si and Al enrichment in the Earth's core may explain the higher-than-chondritic ratios in the shallower, accessible Earth.

## 4. Conclusions

We would like to emphasize that this work does not aim to develop a comprehensive thermodynamic model of Earth's core—for which many other important factors need to be included—but is instead focused on identifying new chemistries for Fe at terrestrial core conditions. For example, Earth's core may contain a considerable budget of light elements such as S, O and H that might modify trace element affinities for the core [19]. However, since the Earth's core consists of predominantly Fe (and Ni) that will largely lower the activity of the light elements, the inclusion of these light elements is unlikely to overturn the chemistry of *p*-block elements in the core that has been examined here. Furthermore, our calculations are performed for crystalline compounds, and are therefore more directly related to Earth's solid inner core. On the other hand, the general trend that the chemical binding with Fe becomes much stronger under pressure can be equally applied to the liquid outer core, because the chemical driving force is irrelevant to the state of the matter.

**Acknowledgements**

M.M. acknowledges the support of National Science Foundation CAREER award (No. 1848141) and the support from American Chemical Society Petroleum Research Fund (No. 59249-UNI6). X.W. was supported by the National Natural Science Foundation of China No.s 11674144 and 11974154, and the Natural Science Foundation of Shandong Province under grant No.s JQ201602 and 2019GGX103023. X.F. acknowledges funding from the China Scholarship Council. S.A.T.R. is grateful for the support of the UK Natural Environment Research Council, through Grant No. NE/P012167/1. J. L. was supported by the Natural Science Foundation of Shandong Province under grant No. ZR2018MA038. Large part of the calculations was performed on NSF-funded XSEDE resources (TG-DMR130005) especially on the Stampede cluster run by Texas Advanced Computing Center.

**References**

[1] W. F. McDonough and S.-S. Sun, Chem. Geol. **120**, 223 (1995).

[2] D. L. Anderson, Proc. Natl. Acad. Sci. **99**, 13966 (2002).

[3] B. J. Wood, M. J. Walter, and J. Wade, Nature **441**, 825 (2006).

[4] E. Anders and N. Grevesse, Geochim. Cosmochim. Acta **53**, 197 (1989).

[5] Z. Wang and H. Becker, Nature **499**, 328 (2013).

[6] K. Lodders and B. Fegley, *The Planetary Scientist's Companion* (Oxford University Press on Demand, 1998).

[7] L. Zhu, H. Liu, C. J. Pickard, G. Zou, and Y. Ma, Nat. Chem. **6**, 644 (2014).

[8] E. Stavrou, Y. Yao, A. F. Goncharov, S. S. Lobanov, J. M. Zaug, H. Liu, E. Greenberg, and V. B. Prakapenka, Phys. Rev. Lett. **120**, 96001 (2018).

[9] X. Du, Z. Wang, H. Wang, T. Iitaka, Y. Pan, H. Wang, and J. S. Tse, ACS Earth Sp. Chem. **2**, 711 (2018).

[10] Q. Hu, D. Y. Kim, W. Yang, L. Yang, Y. Meng, L. Zhang, and H.-K. Mao, Nature **534**, 241 (2016).

[11] J. Botana and M.-S. Miao, Nat. Commun. **5**, 4861 (2014).

[12] J. Feng, R. G. Hennig, N. W. Ashcroft, and R. Hoffmann, Nature **451**, 445 (2008).

[13] W. Zhang, A. R. Oganov, A. F. Goncharov, Q. Zhu, S. E. Boulfelfel, A. O. Lyakhov, E. Stavrou, M. Somayazulu, V. B. Prakapenka, and Z. Konôpková, Science (80-. ). **342**, 1502 LP (2013).

[14] M. S. Miao, Nat. Chem. **5**, 846 (2013).

[15] A. R. Oganov and S. Ono, Nature **430**, 445 (2004).

[16] V. A. Vernikovsky and N. V Sobolev, Russ. Geol. Geophys. **57**, 3 (2016).

[17] K. Lodders, Astrophys. J. **591**, 1220 (2003).

[18] A. F. Goncharov, V. V. Struzhkin, J. A. Montoya, S. Kharlamova, R. Kundargi, J. Siebert, J. Badro, D. Antonangeli, F. J. Ryerson, and W. Mao, Phys. Earth Planet. Inter. **180**, 148 (2010).

[19] J. Badro, A. S. Côté, and J. P. Brodholt, Proc. Natl. Acad. Sci. **111**, 7542 (2014).

[20] A. Jain, S. P. Ong, G. Hautier, W. Chen, W. D. Richards, S. Dacek, S. Cholia, D. Gunter, D. Skinner, G. Ceder, and K. A. Persson, APL Mater. **1**, 011002 (2013).

[21] Y. Wang, J. Lv, L. Zhu, and Y. Ma, Phys. Rev. B **82**, 094116 (2010).

[22] Y. Wang, J. Lv, L. Zhu, and Y. Ma, Comput. Phys. Commun. **183**, 2063 (2012).

[23] J. P. Perdew, K. Burke, and M. Ernzerhof, Phys. Rev. Lett. **77**, 3865 (1996).

[24] P. E. P. Blöchl, Phys. Rev. B **50**, 17953 (1994).

[25] G. Kresse and D. Joubert, Phys. Rev. B **59**, 1758 (1999).

[26] G. Kresse and J. Furthmüller, Phys. Rev. B **54**, 11169 (1996).

[27] H. J. Monkhorst and J. D. Pack, Phys. Rev. B **13**, 5188 (1976).

[28] G. Henkelman, A. Arnaldsson, and H. Jónsson, Comput. Mater. Sci. **36**, 354 (2006).

[29] M. J. Walter, E. Nakamura, R. G. Trønnes, and D. J. Frost, Geochim. Cosmochim. Acta **68**, 4267 (2004).

[30] D. C. Rubie, D. J. Frost, U. Mann, Y. Asahara, F. Nimmo, K. Tsuno, P. Kegler, A. Holzheid, and H. Palme, Earth Planet. Sci. Lett. **301**, 31 (2011).

[31] C. R. M. Jackson, L. B. Ziegler, H. Zhang, M. G. Jackson, and D. R. Stegman, Earth Planet. Sci. Lett. **392**, 154 (2014).

SUPPLEMENTARY INFORMATION

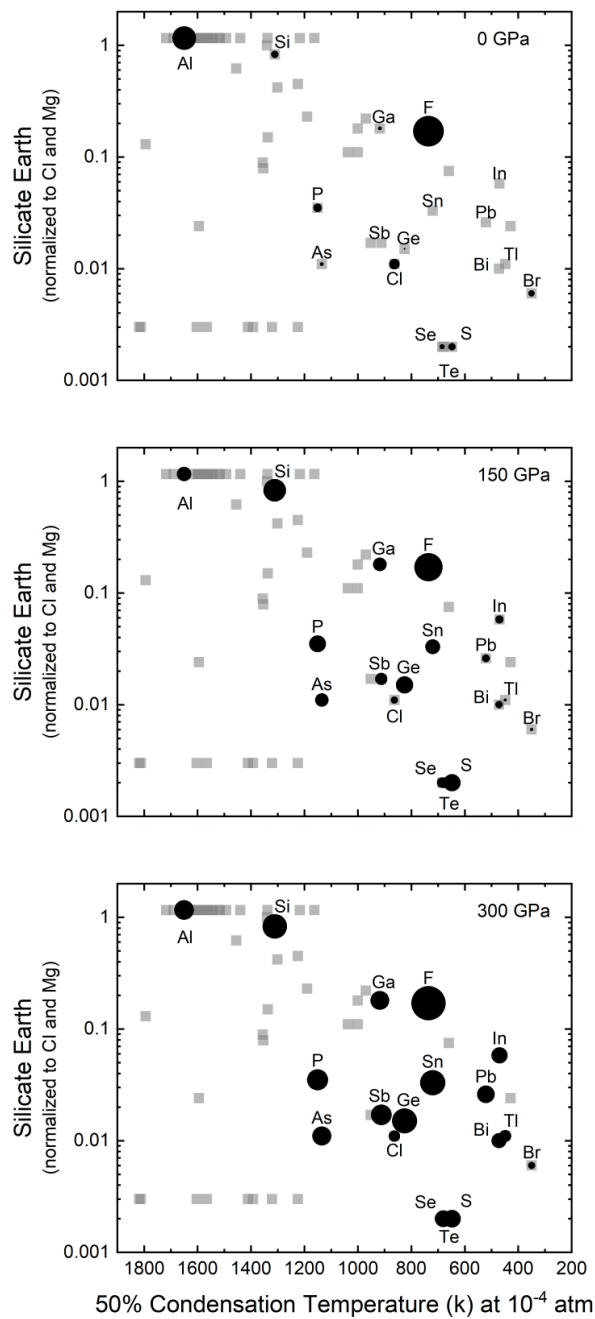

**Fig. S1.** Reproduction of a plot of the abundance of elements in the Silicate Earth (normalized to Mg and CI chondrites) versus their 50% condensation temperature at 10-4 atm. The size of the filled circles represents the binding strength of the element with Fe quantized by the enthalpy of formation per atom (larger symbols represent higher/lower binding strength). Three panels show three different pressures, 0, 150 and 300 GPa.

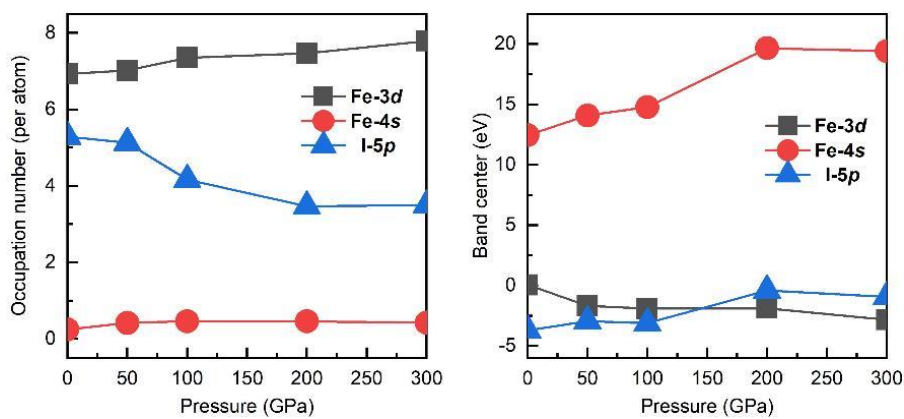

**Fig. S2.** Occupation number and band center as function of pressure in Fe-I compounds. At each pressure, the compound that have the lowest enthalpy of formation per atom is chosen. The occupation numbers are calculated by projecting and integrating the wavefunctions to the atomic orbitals inside the Wigner-Seitz Radius. The band centers are calculated from the density of states (Extended Data Figure 5) using the DOS analysis tool "dosanalyze.pl" in the VTST tool package from UT Austin.

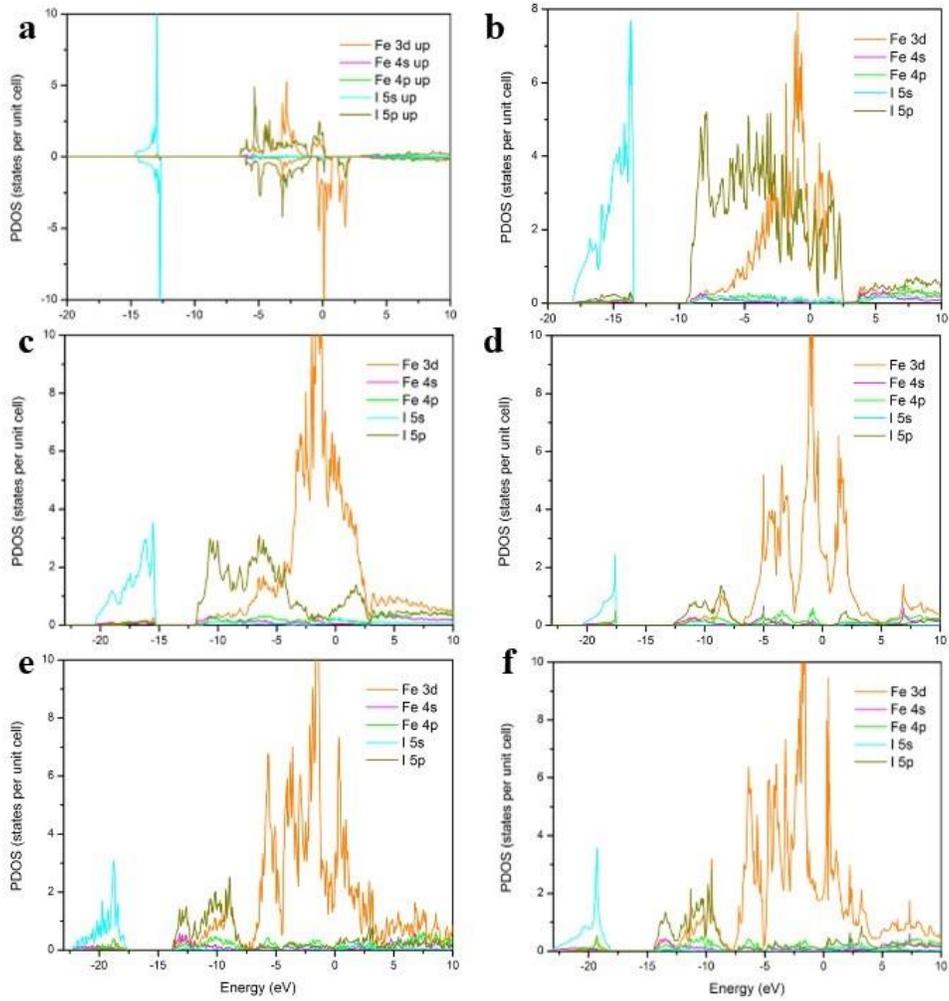

**Fig. S3.** The projected DOS spectra of a, FeI$_2$ with *P*-3*m*1 structure at 0 GPa; b, FeI$_3$ with *R*-3 structure at 0 GPa; c, FeI$_3$ with *R*32 structure at 50 GPa; d, FeI with *P*-1 structure at 100 GPa; e, Fe3I with *Pm*-3*m* structure at 200 GPa; f, Fe2I with *P*6$_3$/*mmc* structure at 400 GPa.

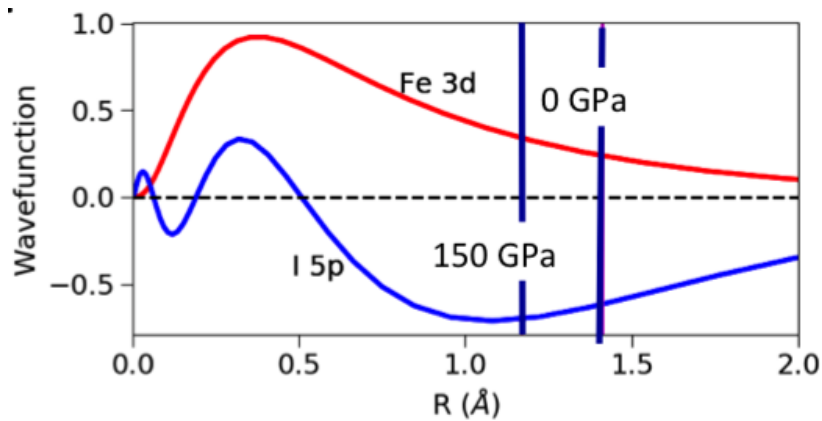

**Fig. S4.** The free-atom wavefunctions of the Iron 3d and Iodine 5p orbitals. The wavefunctions are calculated by use of the Fritz-Harber-Institute (FHI) pseudopotential generation package. The two vertical lines show the radii that are half of the Fe-I distances at 0 GPa and 150 GPa in the Fe-I compounds that have the lowest enthalpy of formation per atom at that pressure. It clearly shows that the compression of the compounds interferes much more significantly with Iodine 5p orbitals than with Fe 3d orbitals. Therefore, the energy levels of Iodine 5p will increase much faster with pressure than that of Iron 3d, which is the driving force of the charge transfer reversal in Fe-I compounds under pressure.

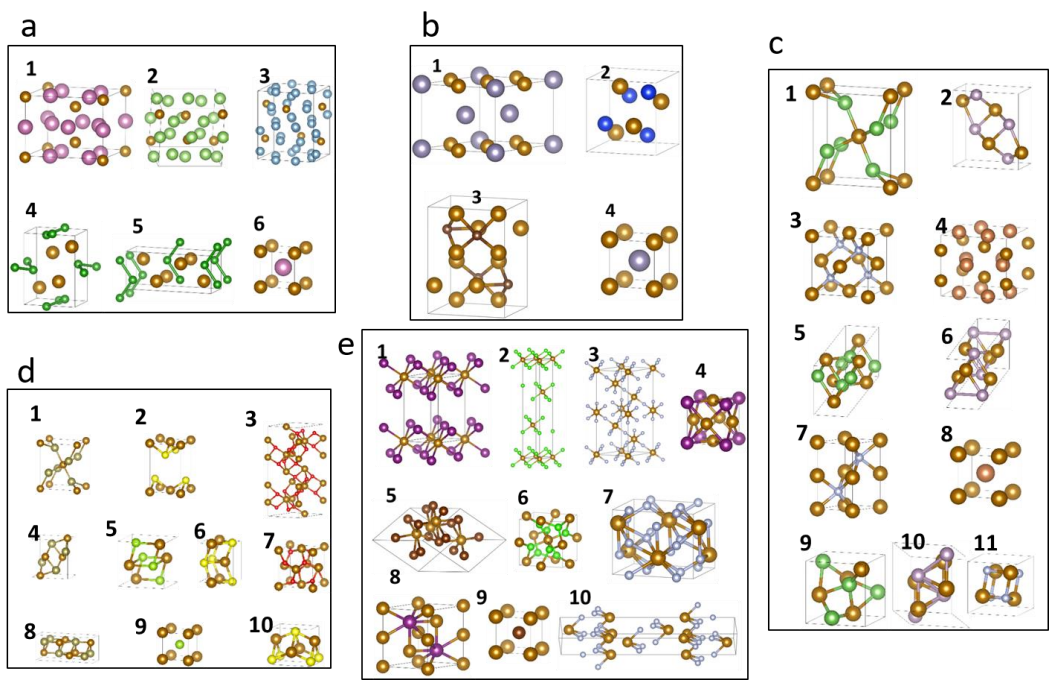

**Fig. S5.** More crystal structures of Fe compounds with p elements, at 0 GPa, 150 GPa and 300 GPa. **a)** 1, FeIn$_3$ with *P4* space group under 0 GPa; 2, FeGa$_3$ with *P4$_2$/mnm* space group under 0 GPa; 3, FeAl$_6$ with *Cmcm* space group under 0 GPa; 4, FeB with *Pnma* space group under 0 GPa; 5, FeB with *Cmcm* space group under 150 GPa or 300 GPa; 6, FeIn, FeGa and FeAl with *Pm-3m* space group under 150 or 300 GPa. **b)** 1, FeSn and FeGe with *P6/mmm* space group under 0 GPa; 2, FeSi with *P2$_1$3* space group under 0 GPa; 3, Fe$_3$C with *Pnma* space group under 0 GPa, 150 GPa or 300 GPa; 4, FeSn, FeGe and FeSi with *Pm-3m* space group under 150 GPa or 300 GPa. **c)** 1, FeSb$_2$ and FeAs$_2$ with *Pnnm* space group under 0 GPa; 2, FeP with *Pnma* space group under 0 GPa; 3, FeN with *F-43m* space group under 0 GPa; 4, FeSb$_2$ with *I4/mcm* space group under 150 GPa; 5, FeAs with *C2/m* space group under 150 GPa; 6, FeP with *C2/m* space group under 150 GPa; 7, FeN with *P6$_3$/mmc* space group under 150 GPa; 8, FeSb with *Pm-3m* space group under 300 GPa; 9, FeAs with *P2$_1$3* space group under 300 GPa; 10, FeP with *P2$_1$/c* space group under 300 GPa; 11, FeN with *P2$_1$3* space group under 300 GPa. **d)** 1, FeTe$_2$ and FeSe$_2$ with *Pnnm* space group under 0 GPa; 2, FeS with *P4/nmm* space group under 0 GPa; 3, Fe$_2$O$_3$ with *R-3c* space group under 0 GPa; 4, FeTe with Pnma space group under 150 GPa; 5, FeSe with *Pbcm* space group under 150 GPa; 6, FeS with *Pmmn* space group under 150 GPa; 7, FeO$_2$ with *Pa-3* space group under 150 GPa or 300 GPa; 8, FeTe with Cccm space group under 300 GPa; 9, FeSe with *Pm-3m* space group under 300 GPa; 10, FeS with *Fmmm* space group under 300 GPa. **e)** 1, FeI$_2$ and FeBr$_2$ with *P-3m1* space group under 0 GPa; 2, FeCl$_2$ with *R-3m* space group under 0 GPa; 3, FeF3 with R-3c space group under 0 GPa; 4, Fe$_3$I with *Pm-3m* space group under 150 GPa; 5, FeBr with *P-1* space group under 150 GPa; 6, FeCl$_2$ with *Pa-3* space group under 150GPa; 7, FeF$_3$ with Cmcm space group under 150 GPa; 8, Fe$_2$I with *P6$_3$/mmc* space group under 300 GPa; 9, FeBr and FeCl with *Pm-3m* space group under 300 GPa; 10, FeF with *R-3m* space group under 300 GPa.

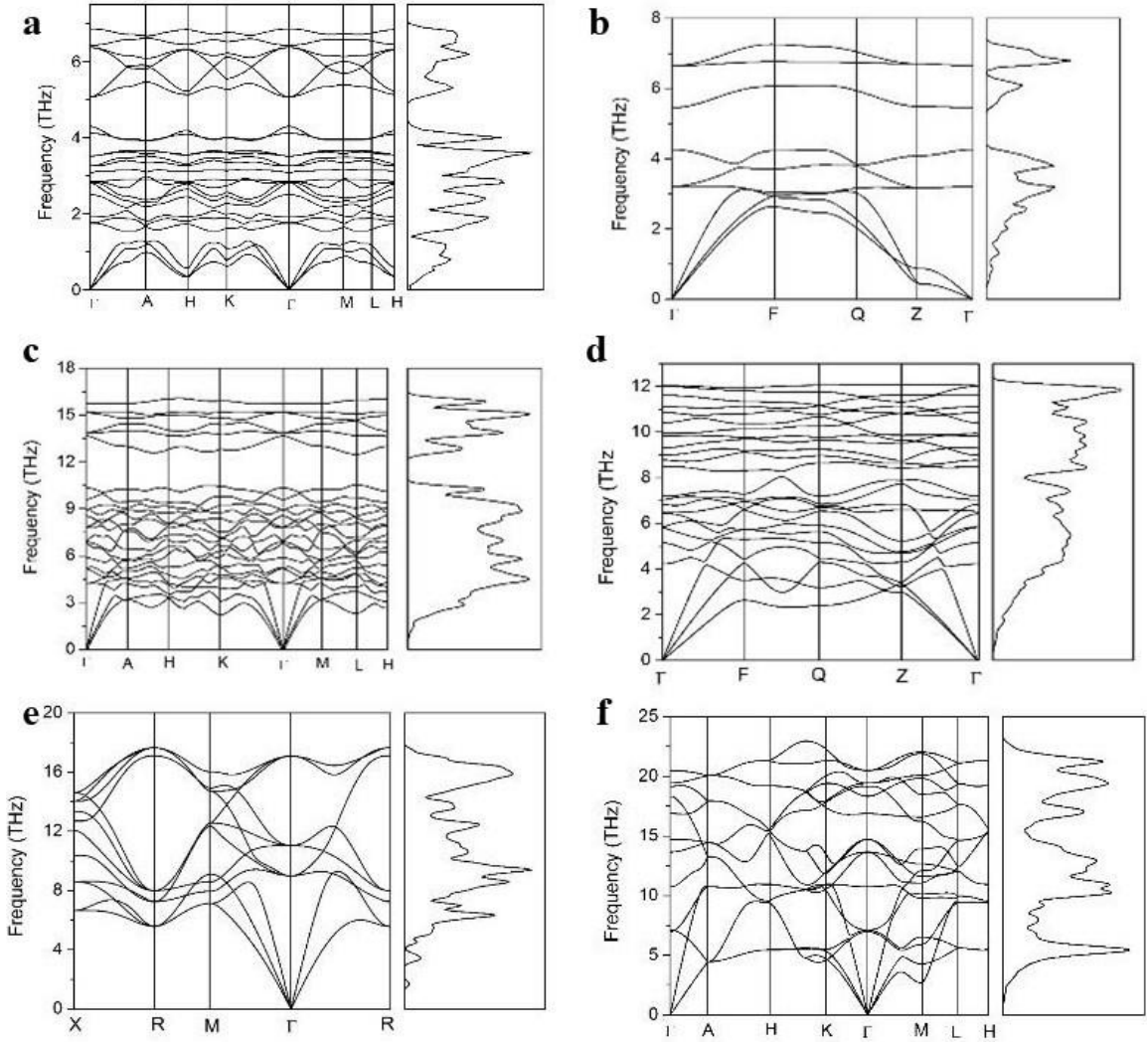

**Fig. S6.** The phonon spectra of FeI compounds shown as example Fe-X compounds for their dynamic stability. **(**The phonon dispersion (left) and density of states (right)). (a)FeI$_2$ with *P*-3*m*1 structure at 0 GPa, (b) FeI$_3$ with *R*-3 structure at 0 GPa, (c) FeI$_3$ with *R*32 structure at 50 GPa, (d) FeI with *P*-1 structure at 100 GPa, (e) Fe$_3$I with *Pm*-3*m* structure at 200 GPa, and (f) Fe$_2$I with *P*6$_3$/*mmc* structure at 400 GPa. The phonon spectra are calculated using finite displacement method by combining the features of phonopy (http://phonopy.sourceforge.net) and VASP programs.

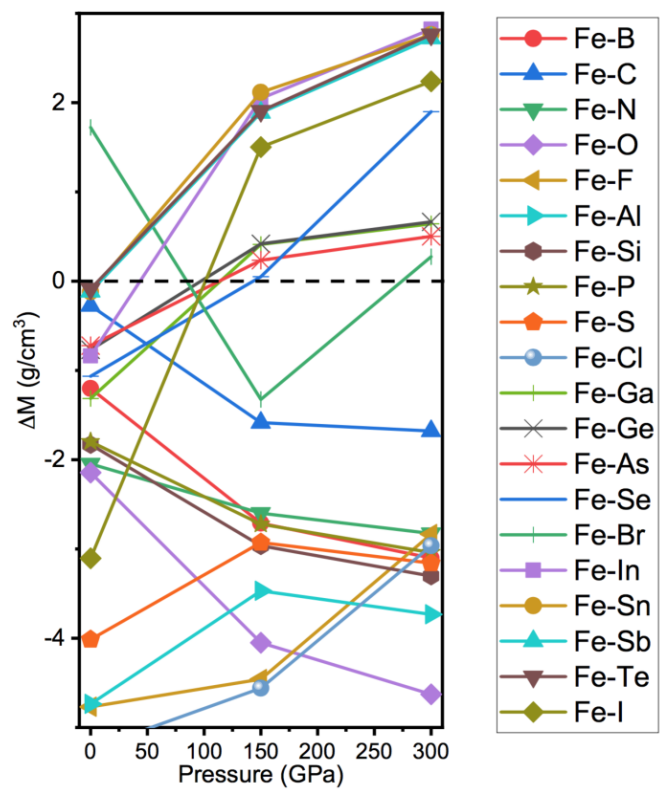

**Fig. S7.** The density difference between Fe$_m$X$_n$ (X= p elements) compounds and pure Fe at 0 GPa, 150 GPa and 300 GPa.

**Discussion 1. P block elements and their significance in Earth science**

This selection of elements includes three major element S (721 K; 50% condensation temperature of 664 K), Si (2628 K; 1310 K) and P (556 K; 1229 K) and a suite of geochemical tracers Ge (3103 K; 883 K), As (886 K; 1065 K), Se (958 K; 697 K), Sn (2543 K; 704 K), Sb (2023 K; 979 K), and Te (1263 K; 709 K) which are critical for understanding planetary accretion and core formation. (The temperatures in parentheses give the 1-bar boiling point and the 50% equilibrium condensation temperature at $10^{-4}$ bar total pressure for solar system abundances of each element, respectively, as a guide to their volatility and reactivity in the condensing solar nebula.) Condensation temperatures are useful to consider when assessing if a particular element is deficient in the silicate mantle relative to chondrites because it suffered volatilization during planetary formation; it is a major task to explore whether such elemental deficiencies relate to elements being "hidden" in the metallic core.

**Discussion 2. Stability of Fe-X (X=p block elements) compounds under pressure**

At ambient conditions, although some $2p$ elements (e.g. C, N, O) can form stable compounds with iron with a $\Delta H_f$ of approximately -0.5 eV/atom, a subset of $3p$ (e.g. Si, P, S; Fe – Al compounds are excepted, and will be discussed separately) and $4p$ (e.g. Ge, As, Se) elements bind only loosely with iron. This general trend of reactivity is markedly changed upon increasing pressure: most of these elements can form stable compounds with iron with a $\Delta H_f$ of at least -1eV/atom. For example, Fe and Te are not likely to form a stable compound (with $\Delta H_f$ = -0.2 eV/atom) at 0 GPa. However, $\Delta H_f$ decreases by 1 eV/atom under high pressure, making the compound FeTe as stable as FeS at 300 GPa, a pressure similar to that in Earth's inner core of 367 GPa).

**Discussion 3. The structure evolution of Fe compounds under pressure**

Both of the two stable Fe-I compounds adopt layered structures at ambient pressure, including FeI$_2$ in MoS$_2$ ($P\bar{3}m1$) and FeI$_3$ in FeBr$_3$ ($R\bar{3}$) structure with lone pairs on iodine anions pointing toward the space between layers (Figs. 3a and 3b). However, at 150 and 300 GPa, iron-rich compounds adopt densely-packed structures, including Fe$_2$I adopting Ni$_2$In structure ($P6_3/mmc$) (Fig. 3c) and Fe$_2$I adopting Cu$_3$Au structure ($Fm\bar{3}m$) (Fig 3d). Correspondingly, the coordination number of iodine increases substantially, to 6 in Fe$_2$I and 12 in Fe$_3$I, and the lone pairs disappear, in good accordance with the CTR under pressure. Similar structure evolution is also found in other compounds containing lone pair electrons at ambient pressure, including Fe-As and Fe-Te compounds. (See Fig. 3e for ambient phase of both $FeAs_2$ and $FeTe_2$, Figs. 3f and 3g and for high pressure phases of FeAs and FeTe, respectively.) In contrast, no lone pair is found in the low-pressure structures of Fe$X$ where $X$ is a group 13 or 14 element. FeSn and FeGe are stable in a highly symmetric $P6/mmm$ structure that contains Fe – Fe inter-metallic bonds (Fig 3h). While increasing pressure induces the large charge transfer to iron, Fe – Fe bonds disappear and the compounds become more ionic. At pressures above 150 GPa, FeSn, FeGe and FeSi transform into the CsCl structure ($Pm\bar{3}m$) (Fig 3i) which is a common structure for AB type ionic compounds when the radius of $A^+$ and $B^-$ ions are similar, as stated by Pauling's first rule. In fact, many Fe$X$ compounds adopt this simple CsCl structure under high pressure, due to the increasing ionic bond character.